*Subject Section*

# GCnet: Using Granger causality to explore the dynamic causality relations among genes associated with intellectual disability in human brain

Lukas Madsen Brandt[1], Katja Nowick [2,*], Jing Qin[1,*]

[1]Department of Mathematics and Computer Sciences, University of Southern Denmark. Campusvej 55, DK-5230, Denmark. [2]Department Institute of Biology – Zoology, Freie Universität Berlin, Germany.

*To whom correspondence should be addressed.

## Abstract

**Motivation:** Intellectual disability (ID) is defined by an IQ under 70, in addition to deficits in two or more adaptive behaviors that affect everyday living. Throughout history, individuals with ID have often been marginalized from society and continue to suffer significantly even in modern times. A varying proportion of ID cases are attributable to genetic causes. Identifying the causal relation among these ID-associated genes and their gene expression pattern during brain development process would gain us a better understanding of the molecular basis of ID.

**Results:** In this paper, we interpret gene expression data collected at different time points during the in vitro brain development process as time series and further introduce Granger causality test to evaluate the dynamic dependence relations among genes. These evaluations are used as input to construct gene expression network and extract the pathological information associated to ID including identifying new genes that can be critically related to the disease. To demonstrate our methods, we provide a priority list of new genes that are most likely associated with Mowat Wilson Syndrome via monitoring the community structure of ZEB2 in our Granger causality network constructed based on the Kutsche dataset (Kutsche, et al., 2018).

**Availability:** https://github.com/LukasMadsenBrandt/gene_analysis_dashboard
**Contact:** qin@imada.sdu.dk (Jing Qin) and katja.nowick@fu-berlin.de (Katja Nowick)
**Supplementary information:** Supplementary data are available at *Bioinformatics* online.

## 1 Introduction

Intellectual disability (ID) is defined by an IQ under 70, in addition to deficits in two or more adaptive behaviors that affect everyday living. It affects about 2-3% of the general population (Daily, et al., 2000; Disease, et al., 2016). Generally, patients with ID have differing degrees of difficulties within three adaptive functional domains: conceptual skills (language, knowledge, and memory), social skills and practical skills. Individuals with very severe ID will depend on caregivers for the duration of their life. Currently there is no cure for ID. Throughout human history individuals with ID have often been marginalized from society and continue to suffer significantly even in modern times.

A varying proportion of ID cases (ranging from 17% to 50%) are attributable to genetic causes (Kaufman, et al., 2010; Moeschler, et al., 2006). A large number of genes are known to be associated with ID (Chiurazzi, et al., 2008; Kaufman, et al., 2010; Lee, et al., 2018; Mirzaa, et al., 2014; Najmabadi, et al., 2011; Oortveld, et al., 2013). Deciphering the gene expression pattern of these ID-associated genes might provide us attractive opportunities a better understanding of the molecular basis of ID, including common pathological patterns in ID, as suggested in (van Bokhoven, 2011).

Towards this end, brain organoids derived from induced pluripotent stem cells (iPSC) have been shown to resemble particular aspect of a developing brain and have been established as a very suitable 3D model to study brain development, especially in primates, where embryonic samples are extremely scarce (Coronel, et al., 2026; Eichmuller and Knoblich, 2022; Fischer, et al., 2019; Heide, et al., 2018; Lancaster and Knoblich, 2014). This system has been used to experimentally characterize the regulatory networks of genes and their dynamic change during the four days differentiation protocol (Kutsche, et al., 2018).

Based on these experiments, advanced co-expression network analysis methods, including weighted topological overlap networks, can be performed to elucidate interactions among genes (Margolin, et al., 2006; Nowick, et al., 2009; Oldham, et al., 2006; Raina, et al., 2023). However, these methods, can not capture changes in gene expressions and interactions among these genes over time. On the other hand, gene expression data are collected at several time points during the organoid developing process, thus interpreting such data as time series is more effective in preserving information.

Granger causality (GC), introduced in (Granger, 1969), has been a powerful notion for characterizing dependence relations between time series in economics and econometrics. Given that the gene expression data during brain organoid development processes are time series, e.g. as shown in Fig.1, Thus, whether one gene's time series has a significant leading impact towards another gene can be evaluated by GC. Naturally, a significant leading impact can be then interpreted as a link between the pair of genes and therefore lead to the concept of a causality gene expression network. This network, once constructed, can serve as a platform to extract pathological information during brain development process based on in vitro data sets, e.g. Kutsche dataset (Kutsche, et al., 2018).

In this paper, we introduce GC to measure the significance of the dynamic association among gene pairs during brain development process and provide critical pathological information associated with ID through downstream network analysis. To demonstrate our methods, we use Mowat Wilson Syndrome as an example and ZEB2 as a stepstone in the network analysis given that mutations in ZEB2 can be a cause this syndrome (Hossain, et al., 2025; Yamada, et al., 2014). In addition, ZEB2 is involved in the formation of the neural tube and neural crest, cortical neurogenesis, hippocampal formation and myelination, and induction of glycogenesis in embryonic and postnatal neocortical progenitors (El-Kasti, et al., 2012; Rogers, et al., 2013; Seuntjens, et al., 2009; Vandewalle, et al., 2009; Weng, et al., 2012). All these facts suggests that ZEB2 functions through a complex gene regulatory network.

This paper is organized as following: data preprocessing and a brief description of GC test are given in Section 2.1 and 2.2, respectively. Furthermore, we describe our disease-oriented network analysis method in Section 2.3 and 2.4. In Section 3, we elucidate our discovery regarding Mowat Wilson Syndrome via ZEB2 and followed up by a discussion of our future perspective in Section 4.

## 2 Methods

### 2.1 Data Preprocessing

Our study is based on Kutsche dataset (Kutsche, et al., 2018). Kutsche dataset contains the expression levels of the 56269 genes in total to provide time-series data on gene expression levels and thus elucidate the dynamic regulatory relationships among genes during the development of the brain. The expression level data is collected from iPSCs as they develop into neurons over 4 days at Day 0, Day 1, Day 2, Day 3, and Day 4. At each collecting point, 7 biological replicas are collected.

Among these 56269 genes, 10365 of these genes are not observed with any positive expression numbers during these 4 days, i.e. only with 0's. Thus, our focus is on the remaining 45904 genes that are expressed. For each of these genes, we summarize multiple replicas collected at the same day into one representative value via distance weighted median and then link consecutive days' representatives together, as shown in Fig. 1. Our default choice of using distance weighted median is due to its robustness against outliers. Alternatively, one could also choose median or mean for this task.

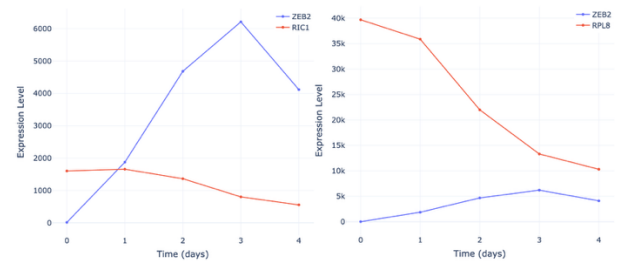

**Fig. 1.** Expression time series of two gene pairs. Based on the Granger causality test, RIC1 causes ZEB2, while ZEB2 cause RPL8 at the default p-threshold level 0.0005.

### 2.2 Granger Causality Test

Granger causality (GC) is a concept formalized by Granger in the context of linear regression models of stochastic processes which is based on the original idea developed by Wiener in 1956. Its intuition is that: one time series is considered has a leading effect on the other if we can better predict the latter time series by incorporating information of the first one.

Apply this concept in our context, we say, 'Gene A causes Gene B' if adding Gene A's past observations can significantly improve the prediction of Gene B, compared to using solely Gene B's own previous observations.

More precisely, given a pair of time series of Gene A, $\{x_t\}$ and Gene B, $\{y_t\}$, respectively for time stamps $t$ ranges from 1 to $T$. The hypothesis test is built based on comparing the following two linear regression models

$$y_t = c_t + \sum_{i=1}^{l} a_i y_{t-i} + \sum_{j=1}^{l} b_j x_{t-j} + \varepsilon_t \quad (1)$$

$$y_t = c_t + \sum_{i=1}^{l} d_i y_{t-i} + \delta_t \quad (2)$$

The null hypothesis is that Gene A does *not* cause Gene B, which can be equivalently interpreted as the Model (2) performs at least as well as Model (1) in the sense of mean squared errors. I.e. the best prediction model of A in the form of (1) satisfies

$$b_1 = b_2 = \cdots = b_l = 0$$

Under this null hypothesis, set residues as $RSS_1 = \sum_{t=1}^{T} \varepsilon_t^2$, $RSS_2 = \sum_{t=1}^{T} \delta_t^2$ and lag $l$, we have the test statistic

$$\frac{(RSS_1 - RSS_2)/l}{\frac{RSS_2}{T - 2l - 1}} \sim F_{l, T-2l-1}$$

This is in practice implemented in the existing Python library *statsmodels*.

Our default p-value threshold is 0.0005, which is determined based on both the computational demand (see Supplement Material (SM) Table 1) and the GC-test noise level for genes. For each gene, to determine its noise level, we conduct GC-tests between this gene and randomly generated white-noise time series. The number of random series that reject the null hypothesis is viewed as the noise level of this gene.

We further construct the (directed) network based on the Granger causal relations among gene pairs. In which, a directed link is drawn from Gene A to Gene B if Gene A causes Gene B. We refer to the resulting network, GC network in the following.

**2.3 Consensus community detection by Louvain Algorithm**
Once the GC network is constructed, to identify new ID-associated genes as described in Section 2.4, community detection is needed. Many algorithms have been developed to uncover the community structure, see review (Lancichinetti and Fortunato, 2009; Li, et al., 2024). We chose Louvain community detection algorithm (Blondel, et al., 2008), due to its efficiency for detecting communities within large networks. Note that Louvain algorithm has a random component due to its random initialization, and different initial conditions can lead to different output community structures.

To tackle this variability, we construct a large sample of community structures derived via independently running Louvain algorithm until a predefined stop criterion is met. The core element of stability criterion is an $N \times N$ matrix which we refer to as co-association matrix, where $N$ is the number of genes. For each pair of genes, say Gene $i$ and Gene $j$, the $(i, j)$-term of this matrix records the frequency of this pair are observed in the same community among different simulations. A higher co-association score indicates a higher likelihood these two genes belong to the same community. The stop criterion to detect a consensus community is, for every 1000 successive increments of independent Louvain runs,

- $\geq 90\%$ of the relative difference within the co-association matrix fell below 5% and
- $\geq 95\%$ of the top frequent genes (5% of the genes in the network) stay the same. Afterwards, the community structure involving these top genes is retrieved as vertex induced subgraph of the original network.

Our adaptive approach is implemented in Python, leveraging several computational libraries: NetworkX is utilized for efficient network management and manipulation. Scikit-learn facilitates hierarchical clustering, specifically through its robust implementation of Agglomerative Clustering. Furthermore, concurrent.futures is used to parallelize the multiple Louvain algorithm runs, significantly reducing computational time and making the analysis of thousands of iterations practical and manageable.

2.4 Identify new ID-associated genes
Recall that Kutsche dataset contains time series for in total 45904 genes which positive expression levels are observed at least twice. Carrying out the entire stream of Granger causality analysis and the follow-up community detection described previously among these 45094 genes would be rather time consuming, at least for moderate computation resources.

Therefore, we introduce the following heuristic routine which is applicable with two additional inputs
a) a pre-existing list of 2310 genes that are known to be associated with microcephaly or intellectual disability in humans according to the human phenotype ontology database (Gargano, et al., 2024) and

b) one hint gene is selected among these 2310 genes and is known to be closely associated to a particular ID syndrome of interest. For example, ZEB2 as our hint gene for the Mowat Wilson Syndrome.

To detect new genes associated with a target ID syndrome, GCnet follow a 2-step procedure:
1) Community detection within 2310 genes: GC-tests among 2310 genes are carried out. GCnet computes the consensus community of the hint gene and identifies the top genes within this community are identified as core genes;
2) Exploration in 45904 genes. GCnet conducts GC-test between core genes and all the 45904 genes available in the Kutcher dataset to identify the significant causal links between core genes and the whole set of genes under default p-value threshold. Add new obtained links to the existing structure of 2310 genes and then deploy the community detection again to obtain an extended list of core genes. Note that, new genes that are not listed in the 2310 genes are identified as new ID syndrome associated genes. Among these genes, their frequencies to share the same community as the hint gene are used as a score function to prioritize their potential to be the specific ID syndrome associated.

## 3 Results
**3.1 Case study: Mowat Wilson Syndrome hinted by ZEB2.**
**Community detection within 2310 genes.** Under the default p-value threshold 0.0005, there are in total 3598 pairs, which involve 2079 genes, are considered as of significant causal associations (SM Table 2). The consensus community detection is applied to detect the community of ZEB2 within this network. Fig 2 visualizes the induced network of the top 5% genes that are most frequently sharing the same community with ZEB2. In total 104 genes (including ZEB2) are identified as core genes (SM Table 3). These core genes are further categorized via Louvain algorithm into 14 sub-communities, as indicated by different colors in Fig 2.

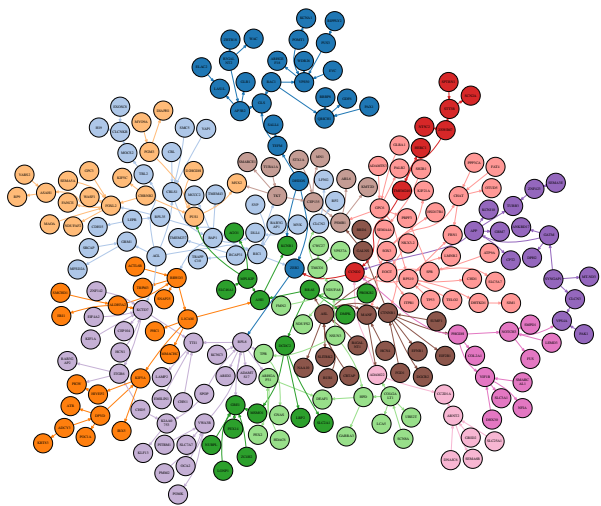

**Fig. 2.** **GC** Network of induced from the 104 (top 5%) most frequent genes which are in the same community with ZEB2 among the 2310 preselected genes. Louvain algorithm is applied to detect sub-communities (indicated by different colors) within this network.

**Exploration within 45904 genes.** Next, under the same p-threshold 0.0005, in total 104 significant links are detected by GCnet between the 104 core genes identified in the previous step and the entire set of genes observed in the Kutcher data set. These new detected significant links (and their associated nodes) are merged with the original network of the 2310 nodes to create a new network of 5178 nodes and 5622 edges (SM Table 3). Within 5278 nodes, 258 genes are identified as core genes, i.e. the genes most frequently within the community of ZEB2 based on community detection simulations. The resulting network induced by these core genes is shown in Fig.3 and listed in SM Table 4. There are 248 new genes that are outside the list of 2310 genes. In particular, there are 53 genes that always share the same community with ZEB2, as illustrated in Fig 4. Note that all 53 genes are outside in the list of 2310 genes.

of the induced network of the core genes as shown in Fig 5 and listed in SM Table 6. In which, 19 sub-communities, indicated by different colors, are detected by Louvain algorithm within this network. Analogous exploration procedure is applied to derive a new network of 9428 nodes and 10946 edges. Further community detection led to 344 core genes that are most frequently sharing the same community with ZEB2. These genes are listed in SM Table 7. There are 330 genes that are outside the list of 2310 genes. In particular, there are 137 genes that always share the same community with ZEB2, as illustrated in Fig 5.

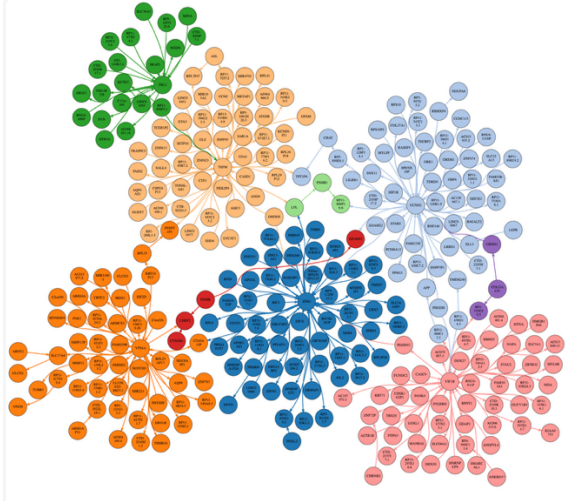

**Fig. 3.** **GC** Network induced from the 258 (top 5%) most frequent genes which are in the same community with ZEB2 after exploration.

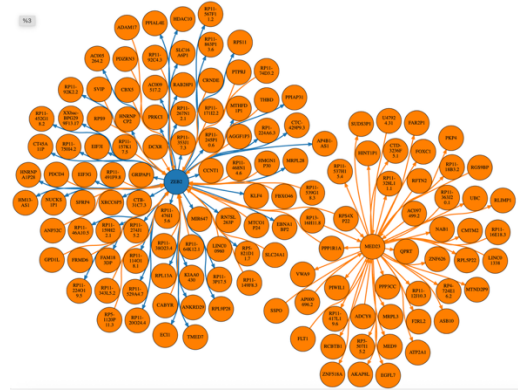

**Fig. 5.** Network induced from 344 core genes which are in the same community with ZEB2 among the 2310 preselected genes under the p-value threshold 0.001.

**Fig. 6.** **GC** Network induced from the 137 genes which are always in the same community with ZEB2 after the exploration under the p-value threshold 0.001.

### Acknowledgements

The authors are alphabetically ordered and with equal contribution.
Thi-Phuong Lee for the list of ID genes.

### Funding

LB was supported by the DDSA travel grant. JQ is supported by the Novo Nordic Recruit Grant 2024.

*Conflict of Interest:* none declared.

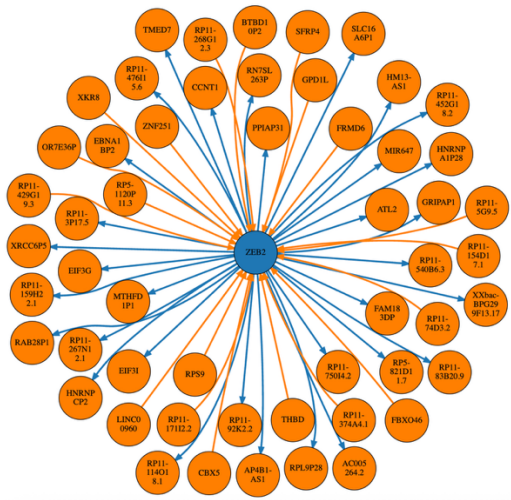

**Fig. 4.** Network induced by the 53 genes that are always within the same community as ZEB2.

## 4 Discussion

After the threshold is relaxed to 0.001, there are in total 6970 pairs, which involve 2277 genes within the 2310 genes, are considered as of significant causal associations. See SM Table 5. Next, the consensus community detection algorithm is deployed to obtain 113 core genes. Furthermore, these 113 genes give rise to an extended list of 450 genes which are the nodes

### References

Blondel, V.D., *et al.* Fast unfolding of communities in large networks. *Journal of Statistical Mechanics: Theory and Experiment* 2008;2008(10):P10008.

Chiurazzi, P., *et al.* XLMR genes: update 2007. *Eur J Hum Genet* 2008;16(4):422-434.

Coronel, R., *et al.* Human cerebral organoids: Complex, versatile, and human-relevant models of neural development and brain diseases. *Neural Regen Res* 2026;21(3):837-854.

**Granger Causality Network for identifying Genes that are potentially associated with Intellectual Disability**

Daily, D.K., Ardinger, H.H. and Holmes, G.E. Identification and evaluation of mental retardation. *Am Fam Physician* 2000;61(4):1059-1067, 1070.

Disease, G.B.D., Injury, I. and Prevalence, C. Global, regional, and national incidence, prevalence, and years lived with disability for 310 diseases and injuries, 1990-2015: a systematic analysis for the Global Burden of Disease Study 2015. *Lancet* 2016;388(10053):1545-1602.

Eichmuller, O.L. and Knoblich, J.A. Human cerebral organoids - a new tool for clinical neurology research. *Nat Rev Neurol* 2022;18(11):661-680.

El-Kasti, M.M., Wells, T. and Carter, D.A. A novel long-range enhancer regulates postnatal expression of Zeb2: implications for Mowat-Wilson syndrome phenotypes. *Hum Mol Genet* 2012;21(26):5429-5442.

Fischer, J., Heide, M. and Huttner, W.B. Genetic Modification of Brain Organoids. *Front Cell Neurosci* 2019;13:558.

Gargano, M.A., *et al.* The Human Phenotype Ontology in 2024: phenotypes around the world. *Nucleic Acids Res* 2024;52(D1):D1333-D1346.

Granger, C.W.J. Investigating Causal Relations by Econometric Models and Cross-Spectral Methods. *Econometrica* 1969;37(3):424-438.

Heide, M., Huttner, W.B. and Mora-Bermudez, F. Brain organoids as models to study human neocortex development and evolution. *Curr Opin Cell Biol* 2018;55:8-16.

Hossain, W.A., *et al.* ZEB2 Gene Pathogenic Variants Across Protein-Coding Regions and Impact on Clinical Manifestations: A Review. *Int J Mol Sci* 2025;26(3).

Kaufman, L., Ayub, M. and Vincent, J.B. The genetic basis of non-syndromic intellectual disability: a review. *J Neurodev Disord* 2010;2(4):182-209.

Kutsche, L.K., *et al.* Combined Experimental and System-Level Analyses Reveal the Complex Regulatory Network of miR-124 during Human Neurogenesis. *Cell Syst* 2018;7(4):438-452 e438.

Lancaster, M.A. and Knoblich, J.A. Organogenesis in a dish: modeling development and disease using organoid technologies. *Science* 2014;345(6194):1247125.

Lancichinetti, A. and Fortunato, S. Community detection algorithms: a comparative analysis. *Phys Rev E Stat Nonlin Soft Matter Phys* 2009;80(5 Pt 2):056117.

Lee, S., *et al.* Gene networks associated with non-syndromic intellectual disability. *J Neurogenet* 2018;32(1):6-14.

Li, J., *et al.* A comprehensive review of community detection in graphs. *Neurocomputing* 2024;600:128169.

Margolin, A.A., *et al.* ARACNE: an algorithm for the reconstruction of gene regulatory networks in a mammalian cellular context. *BMC Bioinformatics* 2006;7 Suppl 1(Suppl 1):S7.

Mirzaa, G.M., *et al.* The Developmental Brain Disorders Database (DBDB): a curated neurogenetics knowledge base with clinical and research applications. *Am J Med Genet A* 2014;164A(6):1503-1511.

Moeschler, J.B., Shevell, M. and American Academy of Pediatrics Committee on, G. Clinical genetic evaluation of the child with mental retardation or developmental delays. *Pediatrics* 2006;117(6):2304-2316.

Najmabadi, H., *et al.* Deep sequencing reveals 50 novel genes for recessive cognitive disorders. *Nature* 2011;478(7367):57-63.

Nowick, K., *et al.* Differences in human and chimpanzee gene expression patterns define an evolving network of transcription factors in brain. *Proc Natl Acad Sci U S A* 2009;106(52):22358-22363.

Oldham, M.C., Horvath, S. and Geschwind, D.H. Conservation and evolution of gene coexpression networks in human and chimpanzee brains. *Proc Natl Acad Sci U S A* 2006;103(47):17973-17978.

Oortveld, M.A., *et al.* Human intellectual disability genes form conserved functional modules in Drosophila. *PLoS Genet* 2013;9(10):e1003911.

Raina, P., *et al.* GeneFriends: gene co-expression databases and tools for humans and model organisms. *Nucleic Acids Res* 2023;51(D1):D145-D158.

Rogers, C.D., Saxena, A. and Bronner, M.E. Sip1 mediates an E-cadherin-to-N-cadherin switch during cranial neural crest EMT. *J Cell Biol* 2013;203(5):835-847.

Seuntjens, E., *et al.* Sip1 regulates sequential fate decisions by feedback signaling from postmitotic neurons to progenitors. *Nat Neurosci* 2009;12(11):1373-1380.

van Bokhoven, H. Genetic and epigenetic networks in intellectual disabilities. *Annu Rev Genet* 2011;45:81-104.

Vandewalle, C., Van Roy, F. and Berx, G. The role of the ZEB family of transcription factors in development and disease. *Cellular and Molecular Life Sciences* 2009;66(5):773-787.

Weng, Q., *et al.* Dual-mode modulation of Smad signaling by Smad-interacting protein Sip1 is required for myelination in the central nervous system. *Neuron* 2012;73(4):713-728.

Yamada, Y., *et al.* The spectrum of ZEB2 mutations causing the Mowat-Wilson syndrome in Japanese populations. *Am J Med Genet A* 2014;164A(8):1899-1908.